\documentstyle[prd,aps]{revtex}

\newcommand{\bea}{\begin{eqnarray}}
\newcommand{\eea}{\end{eqnarray}}

\begin{document}

\draft
\twocolumn[\hsize\textwidth\columnwidth\hsize\csname
@twocolumnfalse\endcsname

\title{The Sachs-Wolfe Effect: Gauge Independence and a General Expression}
\author{Jai-chan Hwang${}^{(a,c)}$ and Hyerim Noh${}^{(b,c)}$}
\address{${}^{(a)}$ Department of Astronomy and Atmospheric Sciences,
                    Kyungpook National University, Taegu, Korea \\
         ${}^{(b)}$ Korea Astronomy Observatory,
                    San 36-1, Whaam-dong, Yusung-gu, Daejon, Korea \\
         ${}^{(c)}$ Max-Planck-Institut f\"ur Astrophysik,
                    Karl-Schwarzschild-Str. 1,
                    85740 Garching bei M\"unchen, Germany}
\date{\today}
\maketitle

\begin{abstract}

We address two points concerning the Sachs-Wolfe effect:
(i) the {\it gauge independence} of the observable temperature anisotropy, and
(ii) a gauge-invariant expression of the effect considering 
{\it the most general situation} of hydrodynamic perturbations.
The first result follows because the gauge transformation of the temperature 
fluctuation at the observation event only contributes to the isotropic 
temperature change which, in practice, is absorbed into the definition of 
the background temperature.
Thus, we proceed without fixing the gauge condition, and
express the Sachs-Wolfe effect using the gauge-invariant variables.

\end{abstract}

\noindent

\vskip2pc]
{\it 1.} 
The excess noise in the radio sky discovered by Penzias and Wilson 
in 1965 \cite{Penzias-Wilson} was immediately recognized as the remnant of the
early hot stage in our universe.
We call it the cosmic microwave background radiation (CMBR).
Soon after its discovery, in a fundamental paper published in 1967 \cite{SW}
Sachs and Wolfe pointed out that the CMBR should show the temperature 
anisotropy caused by photons traveling in the perturbed metric
which is associated with the large-scale structure formation processes based on
the gravitational instability.
The dipole and higher multipole anisotropies are now discovered, \cite{obs}.
There have been many studies of the Sachs-Wolfe effect in the literature, 
\cite{SW-synchronous,SW-zero-shear,Abbott,SW-GI,Kin}.
We notice, however, the two points mentioned in the abstract 
may still deserve addressing.
Despite its trivial nature, we found the first point has not always been well 
understood by workers in the field.
We explain it below Eq. (\ref{T-obs}) and below Eq. (\ref{deltaT-2}).
The general expression mentioned in the second point is in Eq. (\ref{deltaT-4}) 
which is the main result in this {\it Report}.
Section {\it 2.} explains our notation and strategy.

\vskip .5cm
{\it 2.}
As the metric we consider a spatially homogeneous and isotropic
one with the most general spacetime dependent perturbations 
(we set $c \equiv 1$)
\bea
   ds^2 
   &=& - \; a^2 ( 1 + 2 \alpha ) d \eta^2 
       - 2 a^2 ( \beta_{,\alpha} + b Y^{(v)}_\alpha ) d \eta d x^\alpha
   \nonumber \\
   & & + \; a^2 \Big[ g^{(3)}_{\alpha\beta} ( 1 + 2 \varphi ) 
   \nonumber \\
   & & \qquad \quad
       + \; 2 \Big( \gamma_{,\alpha |\beta} + c Y^{(v)}_{(\alpha|\beta)} 
       + C^{(t)}_{\alpha\beta} \Big) \Big] d x^\alpha d x^\beta.
\eea
The four-velocity of the fluid is:
$u^0 \equiv a^{-1} \left( 1 - \alpha \right)$ and
$u^\alpha \equiv a^{-1} ( - k^{-1} v^{(s)|\alpha} + v^{(v)} Y^{(v)\alpha} )$.
$Y^{(v)}_\alpha$ and $C^{(t)}_{\alpha\beta}$ are based on 
$g^{(3)}_{\alpha\beta}$, and a vertical bar indicates the covariant derivative 
based on $g^{(3)}_{\alpha\beta}$.
$Y^{(v)}_\alpha$ is a (transverse) vector harmonic function, 
\cite{Bardeen,vorticity}.
The (tranverse-tracefree) tensor-type perturbation $C^{(t)}_{\alpha\beta}$ 
is invariant under the gauge transformation, and we can construct the 
gauge-invariant combinations for the vector-type perturbation \cite{Bardeen}: 
$v^{(v)} - b \equiv v_\omega$, 
$v^{(v)} + c^\prime  \equiv v_\sigma$, and $v_\sigma - v_\omega \equiv \Psi$,
where a prime denotes the time derivative based on $\eta$.
Due to the spatial homogeneity and isotropy in the background spacetime
the three perturbation types decouple and evolve independently to the 
linear order.

The scalar-type variables depend on the gauge transformation.
Our strategy concerning the gauge is to use the several available
gauge conditions as an advantage for handling problems.
A certain gauge condition is suitable for handling a certain aspect of
the problem. 
But, usually we do not know which gauge is the suitable one {\it a priori}.
Thus, it is desirable to design the equations so that we can easily impose
the fundamental gauge conditions, \cite{Bardeen-1988,PRW}.
We call it the {\it gauge-ready} approach, and the relativistic perturbation 
equations in the gauge-ready form needed in this work can be found in 
\cite{Newtonian}.

Under the gauge transformation $\tilde x^a = x^a + \xi^a$, the variables
transform as ($\xi^t \equiv a \xi^\eta$), \cite{PRW}:
\bea
   & & \tilde \varphi = \varphi - H \xi^t, \;\;
       \tilde \chi = \chi - \xi^t, \;\;
       \tilde v = v - {k \over a} \xi^t, \;\;
       \tilde \alpha = \alpha - \dot \xi^t, 
   \nonumber \\
   & & \tilde \delta = \delta + 3 \left( 1 + {\rm w} \right) H \xi^t, \;\;
       \delta \tilde T = \delta T + H T \xi^t,
   \label{GT}
\eea
where $H \equiv \dot a/a$ and an overdot denotes the derivative with
respect to $t$ ($dt \equiv a d\eta$); ${\rm w} \equiv p/\mu$ where 
$\mu$ and $p$ are the energy density and the pressure, respectively.
$\varphi$, $\chi \equiv a ( \beta + \gamma^\prime )$, 
$v \equiv ( v^{(s)} + k \beta )$, 
and $\delta \equiv \delta \mu / \mu$
are perturbed parts of the three-space curvature, shear, velocity, 
and relative density variables, respectively;
all these variables are spatially gauge-invariant.
{}For the temperature $T({\bf x}, t)$, we decompose it into the background
and perturbed parts as
\bea
   T ({\bf x}, t) = \bar T (t) + \delta T ({\bf x}, t),
   \label{delta-T}
\eea
where an overbar indicates a quantity to the background order; 
we neglect it unless necessary.
If we regard $T$ as a scalar quantity, the perturbed part changes as
$\delta \tilde T = \delta T - \dot T \xi^t$, and
considering $\bar T \propto a^{-1}$, we have Eq. (\ref{GT});
Some of the fundamental gauge conditions we can recognize in Eq. (\ref{GT}) 
are:
the uniform-curvature gauge ($\varphi \equiv 0$), 
the zero-shear gauge ($\chi \equiv 0$),
the comoving gauge ($v \equiv 0$), 
the uniform-density gauge ($\delta \equiv 0$), and
the uniform-temperature gauge ($\delta T \equiv 0$).
Each one of these gauge conditions
fixes the temporal gauge transformation property completely (i.e., $\xi^t = 0$),
and, thus, each variable in these gauge conditions
is equivalent to a corresponding gauge-invariant combination.
The synchronous gauge imposes $\alpha = 0$ and fails to fix the
gauge mode completely; i.e., we still have $\xi^t = \xi^t ({\bf x})$.

We proposed to write the gauge-invariant variables as:
\bea
   & & \delta_v \equiv \delta + 3 (1 + {\rm w}) {aH \over k} v, \;\;
       \varphi_\chi \equiv \varphi - H \chi, \;\;
       \alpha_\chi \equiv \alpha - \dot \chi, 
   \nonumber \\
   & & v_\chi \equiv v - {k \over a} \chi, \;\;
       \varphi_v \equiv \varphi - {a H \over k} v, \;\; 
       {\rm etc}.
   \label{GI}
\eea
$\delta_v$ becomes $\delta$ in the comoving gauge ($v \equiv 0$), etc.
In this manner, using Eq. (\ref{GT}), we can systematically construct 
the corresponding 
gauge-invariant combination for any variable based on a gauge condition which
fixes the temporal gauge transformation property completely.
A given variable evaluated in different gauges can be considered as different
variables, and they show different behaviors in general.

The background universe is described by:
\bea
   H^2 = {8 \pi G \over 3} \mu - {K \over a^2} + {\Lambda \over 3}, \quad
       \dot \mu = - 3 H \left( \mu + p \right), 
   \label{BG-eqs}
\eea
where $K$ and $\Lambda$ are the three-space curvature and the cosmological 
constant, respectively.
Later, it is convenient to have the following equations,
derived in \cite{Newtonian}:
\bea
   & & {k^2 - 3 K \over a^2} \varphi_\chi = 4 \pi G \mu \delta_v,
   \label{Poisson} \\
   & & \dot \varphi_\chi + H \varphi_\chi
       = - 4 \pi G \left( \mu + p \right) {a \over k} v_\chi - 8 \pi G H \sigma,
   \label{v-varphi} \\
   & &  \alpha_\chi = - \varphi_\chi - 8 \pi G \sigma,
   \label{alpha-varphi}
\eea
where $\sigma ({\bf x}, t)$ indicates the anisotropic pressure.

\vskip .5cm
{\it 3.}
The CMBR has a blackbody distribution and the photons are redshifted 
during their travel from last scattering to the observer. 
After the last scattering, the photons are effectively collision-free
and non-self-gravitating, thus follow the geodesic path in the given
(perturbed) metric.
The null vector tangent to the geodesic $x^a (\lambda)$ with an affine 
parameter $\lambda$ is $k^a ={dx^a / d\lambda}$.
We define the null energy-momentum four-vector $k^a$ to the perturbed order as:
$k^0 \equiv a^{-1}( \bar \nu + \delta \nu )$ and
$k^\alpha \equiv -\bar\nu a^{-1} (\bar e^\alpha + \delta e^\alpha )$.
The temperatures of the CMBR at two different points ($O$ and $E$) along 
a single null-geodesic ray in a given observational direction is, \cite{SW},
\bea
   {T_O \over T_E} = {(k^a u_a)_O \over (k^b u_b )_E},
   \label{T}
\eea
where $O$ is the observed event here and now, and $E$ is the emitted event 
at the intersection of the ray and the last scattering surface.
$u_a$ at $O$ and $E$ are the local four-velocities of the observer
and the emitter, respectively.
In the large angular scale we are considering 
(larger than the horizon size at the last scattering era which subtends 
about $2\sqrt{\Omega_0}$ degree by an observer today) 
the detailed dynamics at last scattering is not important.
The physical processes of last scattering are important in the small 
angular scale where we need to solve the Boltzmann equations for the 
photon distribution function, \cite{Kin}.

The observed temperature along the single ray may depend on the 
location of the observer ${\bf x}_O$ (cosmic variance), and the direction 
of the observed ray ${\bf e}_O$.
Similarly as in Eq. (\ref{delta-T}) we may decompose the {\it observed} 
temperature along the single ray into the background and perturbed parts as 
\bea
   T ({\bf x}_O, t_O; {\bf e}_O) = \bar T ({\bf x}_O, t_O) 
       + \delta T ({\bf x}_O, t_O; {\bf e}_O).
   \label{T-obs}
\eea
Although we used similar notations in Eqs. (\ref{delta-T},\ref{T-obs}),
it is desirable to notice the difference:
Eq. (\ref{delta-T}) decomposed the temperature at spacetime points, 
whereas Eq. (\ref{T-obs}) decomposed the observed temperature
along different directions ${\bf e}_O$ at observer's location ${\bf x}_O$.
Up to this point, the decomposition in Eq. (\ref{T-obs}) still has
arbitrariness as the one in Eq. (\ref{delta-T}).
In the observations, however, we often take the background temperature as 
an averaged temperature all around the sky at the observer's location, i.e., 
$\bar T ({\bf x}_O, t_O) \equiv \langle T ({\bf x}_O, t_O; {\bf e}_O) 
\rangle_{{\bf e}_O}$.
In this way the arbitrariness is fixed, and the remaining 
$\delta T |_O$ over the sky apparently coincides with 
the angular variation of observed temperature. 
Thus, $\delta T |_O$ should be independent of the gauge condition
(imposed at the observer's spacetime position).
Let us explain this last point below.
In the temporally evolving background, $\bar T = \bar T(t)$,
$\delta T$ is a gauge dependent quantity.
The gauge dependence of $\delta T$ should be considered in handling
fluctuations at the last scattering era $E$.
However, for $\delta T$ evaluated at the observation event $O$,
the effect of the gauge transformation $H \xi^t ({\bf x}, t)$ evaluated at 
$O$ will show no angular dependence, thus can be {\it absorbed} into 
our definition of the background temperature, and is irrelevant for 
the temperature anisotropy;
thus, the observable temperature {\it anisotropy} is a concept 
{\it independent of the gauge condition} used \cite{cosmic-variance}.
Equivalently, since $H \xi^t |_O$ terms cancel, the difference of observed 
temperatures in two different directions is gauge-invariant.

Perturbation analyses of the null equation ($k^a k_a = 0$), 
the geodesic equation ($k^a_{\;\; ;b} k^b = 0$), and Eq. (\ref{T}) provide 
the equations we need.
To the background order, we have: $\bar T \propto \bar \nu \propto a^{-1}$,
$\bar e^\alpha \bar e_\alpha = 1$,
and $\bar e^{\alpha\prime} = \bar e^\alpha_{\;\;|\beta} \bar e^\beta$.
To the perturbed order, we have
[for convenience, we consider the contributions from three perturbation types 
separately as 
$\delta T |_O = \delta T^{(s)} |_O + \delta T^{(v)} |_O + \delta T^{(t)} |_O$]:
\bea
   & & {\delta T^{(s)}\over T} \Big|_O
       = {\delta T\over T} \Big|_E
       -{1\over k} v_{,\alpha} e^\alpha \Big|^O_E
   \nonumber \\
   & & \qquad \qquad \; \; \;
       +\int^O_E \Big[ - \varphi^\prime +\alpha_{,\alpha} e^\alpha 
       - {1 \over a} \chi_{,\alpha|\beta} e^\alpha e^\beta \Big] dy,
   \label{SW-scalar} \\
   & & {\delta T^{(v)}\over T} \Big|_O
       = v_\omega Y^{(v)}_\alpha e^\alpha \Big|^O_E
       - \int^O_E \Psi Y^{(v)}_{(\alpha|\beta)} e^\alpha e^\beta dy,
   \label{SW-vector} \\
   & & {\delta T^{(t)}\over T} \Big|_O
       = - \int^O_E C^{(t)\prime}_{\alpha\beta} e^\alpha e^\beta dy,
   \label{SW-tensor}
\eea
where ${d / d y} \equiv {\partial / \partial \eta}
- \bar e^\alpha {\partial / \partial x^\alpha}$
[thus, the integral is along the ray's null-geodesic path]. 
The temperature fluctuation in the last scattering era, 
${\delta T/T} |_E$, contributes to the scalar-type perturbation.
Equations (\ref{SW-vector},\ref{SW-tensor}) are apparently gauge-invariant.
Equation (\ref{SW-scalar}) is written in a gauge ready form,
so that we can impose any gauge condition we want.
Each term on the RHS of Eq. (\ref{SW-scalar}) depends on the temporal gauge 
transformation and the gauge invariance of the terms altogether is not obvious; 
using Eq. (\ref{GT}) we can show that the RHS alone is not gauge-invariant
[apparently, the LHS is not also gauge-invariant, so that the overall equation
is gauge-invariant]. 
Shortly, we will see that the observable contributions to anisotropy 
can be expressed in terms of gauge-invariant variables.

\vskip .5cm
{\it 4.}
Now, we concentrate on the scalar-type contribution in Eq. (\ref{SW-scalar}).
In hydrodynamic perturbation based on Einstein gravity, it is known that
only certain variables in certain gauge conditions correctly reproduce
the Newtonian behaviors in the pressureless limit:
the density perturbation variable in the comoving gauge ($\delta_v$), and 
the perturbed potential and the perturbed velocity
variables in the zero shear gauge ($\varphi_\chi$ and $v_\chi$) show the 
correct behavior of the corresponding Newtonian ones, \cite{Lifshitz,Bardeen}.
These correspondences apply in {\it general scales} (including the superhorizon
scale) considering the general $K$ and $\Lambda$, \cite{Newtonian}:
$v_\chi \leftrightarrow \delta v$ and
$- \varphi_\chi \leftrightarrow \delta \Phi$
where $\delta v$ and $\delta \Phi$ are 
the Newtonian velocity and potential fluctuations, respectively.

Using these variables Eq. (\ref{SW-scalar}) can be written in a more 
suggestive form
\bea
   {\delta T^{(s)}\over T} \Big|_O
   &=& {\delta T \over T} \Big|_E
       - {1 \over k} v_{\chi,\alpha} e^\alpha \Big|^O_E
       - \left( \alpha_\chi + H \chi \right) \Big|^O_E 
   \nonumber \\
   & & + \int^O_E \left( \alpha_\chi - \varphi_\chi \right)^\prime dy.
   \label{deltaT-2}
\eea
The gauge dependent terms on the RHS are identified:
the first and $H\chi$ terms are gauge dependent.
Since the $- (\alpha_\chi + H \chi)$ term {\it evaluated at} $O$ (here and now)
does not show the angular dependence, it can be {\it absorbed} into 
the definition of the background temperature; 
this point was noted in \cite{Abbott}. 
The combination of remaining two gauge dependent variables,
$(\delta T/T + H \chi)|_E$, is a gauge-invariant 
combination $\delta T_\chi/T|_E$.
As a matter of fact, by moving $- (\alpha_\chi + H \chi)|_O$ to the LHS
we can make a gauge-invariant form $(\delta T^{(s)}_\chi/T + \alpha_\chi)|_O$.
However, since the added terms only contribute to the isotropic temperature 
changes those do not contribute to the observed angular variation of
temperature in Eq. (\ref{T-obs}) ({\it with} $\bar T$ defined as the
all-sky average); equivalently, the variation of the observed temperature
with directions is gauge-invariant.
Similarly, one can evaluate Eq. (\ref{deltaT-2}) in any gauge condition 
with the same `observable' anisotropy.
In this sense the {\it observable temperature anisotropy} on the LHS 
of Eq. (\ref{deltaT-2}) is {\it gauge-independent}. 

Absorbing the isotropic contributions to $\bar T({\bf x}_O, t_O)$, we have
\bea
   & & {\delta T^{(s)}\over T} \Big|_O
       = - {1 \over k} v_{\chi,\alpha} e^\alpha \Big|_O
       + {1 \over k} v_{\chi,\alpha} e^\alpha \Big|_E
   \nonumber \\
   & & \qquad \qquad
       + \Big( \alpha_\chi + {\delta T_\chi \over T} \Big) \Big|_E
       + \int^O_E \left( \alpha_\chi - \varphi_\chi \right)^\prime dy.
   \label{deltaT-3}
\eea
The RHS is apparently gauge-invariant.
In the literature, the four terms on the RHS are often called:
the Doppler effect due to the observer's movement, the Doppler effect due to 
the movement of the photon-emitting plasma along the line-of-sight,
the Sachs-Wolfe (SW) effect, and the integrated Sachs-Wolfe (ISW) effect,
respectively, \cite{caution}.

Now, we re-express the SW and the ISW terms using $\varphi_\chi$
which has the close analogy with the Newtonian gravitational potential. 
In order to relate the temperature fluctuation with the coexisting 
matter at $E$, we take an {\it ansatz}
\bea
   {\delta T \over T} \Big|_E \equiv 
       \Big[ {\delta \over 3 (1 + {\rm w})} + e_T \Big] \Big|_E,
   \label{entropic}
\eea
where $e_T ({\bf x}, t)$ is apparently gauge-invariant and can be regarded 
as the deviation of the temperature fluctuation from the adiabaticity with 
the coexisting matter fluctuation; we may call it the 
{\it entropic temperature fluctuation} \cite{e_T}.
By considering $e_T$ we can handle the effects from the multi-component
hydrodynamic situation, \cite{isocurvature}.

Using Eqs. (\ref{BG-eqs}-\ref{alpha-varphi}) we can express 
the SW and ISW terms in Eq. (\ref{deltaT-3}) using $\varphi_\chi$
\bea
   & & {\delta T^{(s,{\rm SW},{\rm ISW})}\over T} \Big|_O
       = \Big\{ \Big[ -1 + {H^2 \over 4 \pi G (\mu + p)} \Big]
       \left( \varphi_\chi + 8 \pi G \sigma \right)
   \nonumber \\
   & & \qquad
       + {H^2 \over 4 \pi G (\mu + p)} \Big( {\dot \varphi_\chi \over H}
       + {k^2 - 3 K \over 3 a^2 H^2} \varphi_\chi \Big) + e_T 
       \Big\} \Big|_E
   \nonumber \\
   & & \qquad
       - 2 \int^O_E \left( \varphi_\chi + 4 \pi G \sigma \right)^\prime dy.
   \label{deltaT-4}
\eea
In this form, we considered the general $K$, $\Lambda$, and $p(\mu)$ 
in the background, and the general $e ({\bf x}, t)$ (the entropic pressure), 
$\sigma$, and $e_T$ in the perturbation.
In an ideal fluid (thus, $e = 0 = \sigma$), the general super-sound-horizon 
scale solution for $\varphi_\chi$ is presented in \cite{Newtonian}
\bea
   \varphi_\chi ({\bf x}, t)
       = 4 \pi G C ({\bf x}) {H \over a} \int_0^t {a (\mu + p) \over H^2} dt
       + {H \over a} d ({\bf x}),
   \label{varphi-chi-sol}
\eea
where $C({\bf x})$ and $d({\bf x})$ are integration constants indicating the
relatively growing and decaying modes, respectively.
Remarkably, this solution is {\it valid} on scales larger than Jeans scale
for the general $K$, $\Lambda$, and generally time-varying $p(\mu)$.
In the near flat case (thus, ignoring $K$ terms), we have a powerful conserved 
quantity in the super-sound-horizon scale: 
$\varphi_v ({\bf x}, t) = C ({\bf x})$, with the vanishing leading 
decaying mode.
The structural seed originated from the quantum fluctuation during the
inflation era provides the initial condition for $C({\bf x})$ and it is 
conserved during the super-sound-horizon scale evolution
independently of changing equation of state, changing gravity theories,
and the horizon crossing, \cite{Newtonian,GGT}.

{}For $K = 0 = \Lambda$ and ${\rm w} = {\rm constant}$, the growing mode 
of $\varphi_\chi$ in Eq. (\ref{varphi-chi-sol}) remains constant
(we ignored $e$ and $\sigma$).
Thus, ignoring the decaying mode, we have
\bea
   {\delta T^{(s,{\rm SW})}\over T} \Big|_O
   &=& \Big\{ \Big[ - {1 + 3 {\rm w} \over 3(1 + {\rm w})} 
       + {2 \over 9(1 + {\rm w})} \Big( {k \over a H} \Big)^2 \Big] \varphi_\chi
   \nonumber \\
   & & \quad 
       + \; e_T \Big\} \Big|_E,
   \label{deltaT-6}
\eea
and the ISW term vanishes.
The large observed angular scale corresponds to the superhorizon scale 
at the time of last scattering, and the effect from $(k/aH)^2$ term
becomes subdominant.
Thus, in the large angular scale, assuming the pressureless era at $E$,
and ignoring $e_T|_E$, we finally have
\bea
   {\delta T^{(s,{\rm SW})}\over T} \Big|_O
       = - {1 \over 3} \varphi_{\chi} \big|_E
       = {1 \over 3} \delta \Phi \big|_E,
   \label{dT-1/3}
\eea
which is the commonly quoted result derived in \cite{SW}.
Notice, however, the various levels of assumptions used to have 
Eq. (\ref{dT-1/3}): we assumed, a single component, pressureless ($p=0$),
adiabatic ($e_T = 0$), ideal fluid ($e = 0 = \sigma$), with $K = 0 = \Lambda$,
and vanishing transient mode at $E$ for the SW term, and along the ray's 
path from $E$ to $O$ for the ISW term.

\vskip .5cm
{\it 5.}
Equation (\ref{deltaT-4}) expresses the SW and the ISW effects
in the very general situation, \cite{comment:C-L}.
Besides this, we also have two Doppler terms in Eq. (\ref{deltaT-3}) 
and the vector and tensor contributions in 
Eqs. (\ref{SW-vector},\ref{SW-tensor}). 
These altogether contribute to the observed temperature anisotropy, \cite{SW}.
Attempts to explain the result in Eq. (\ref{dT-1/3}) in pedagogic ways, 
e.g., \cite{pedagogic}, usually involve gauge dependent interpretations
\cite{interpretation} with limited implications, and should be read
with due caution.
Many works in the literature start by fixing a certain gauge condition 
\cite{SW,SW-synchronous,SW-zero-shear} or by using combinations 
of the gauge-invariant variables \cite{Abbott,SW-GI,C-L}.
The final results for the observed temperature anisotropy are bound to be 
the same as ours, because, as we have shown, the concept is 
{\it observationally gauge-independent}.

\vskip .5cm
We thank Profs. K. Subramanian and A. M\'esz\'aros for useful discussions,
and Prof. S. D. M. White for careful comments and invitation to MPA.
We wish to acknowledge the financial support of the Korea Research Foundation.
HN was supported by the DFG fellowship (Germany) and the KOSEF (Korea).



\begin{thebibliography}{99}
\bibitem{cosmic-variance}
         In the real universe the sky averaged temperature may depend 
         on the observer's location, and contributes to the cosmic variance.
\bibitem{caution}
         One should remember that these are just names used for convenience.
         It is noted that in a time varying gravitational field we cannot
         distinguish the gravitational shift from the Doppler shift 
         in an invariant way, \cite{SW}.
\bibitem{e_T}
         Although Eq. (\ref{entropic}) is motivated by the fact that 
         $\bar T \propto a^{-1}$, and $\bar \mu \propto a^{-3 ( 1 + {\rm w})}$ 
         for ${\rm w} = {\rm constant}$, we can still regard ${\rm w}$ as the
         generally time varying one.
\bibitem{isocurvature}
         In the multi-component situation we regard the fluid quantities
         as the collective ones.
         In the two component situation with dust (or dark matter, $d$) and
         radiation ($r$), it is convenient to introduce 
         $S \equiv \delta_d - {3 \over 4} \delta_r$ which indicates the entropy
         perturbation.
         If we take $\delta \mu = \delta \mu_d + \delta \mu_r = 0$ as the 
         condition for an isocurvature mode, we can show
         $e_T = - S/( 3 + 4 \mu_r/\mu_d)$.
\bibitem{comment:C-L} 
         A similar expression using the covariant approach has been recently
         derived in Eq. (3.17) of \cite{C-L}.
\bibitem{interpretation}
         In the literature we find many analyses based on either 
         the synchronous gauge ($\alpha \equiv 0$) \cite{SW,SW-synchronous} 
         or the zero-shear gauge ($\chi \equiv 0$) \cite{SW-zero-shear}.  
         The often mentioned `gauge dependent' wisdoms are the following 
         [we assume the same conditions used to get Eq. (\ref{dT-1/3})]:
         (i) In the zero-shear gauge, using 
         Eqs. (\ref{Poisson}-\ref{alpha-varphi}), we can show
         $\delta T_\chi/T = {2 \over 3} \varphi_\chi$, thus 
         after compensation with $\alpha_\chi (= - \varphi_\chi)$ in
         Eq. (\ref{deltaT-3}) we have Eq. (\ref{dT-1/3}).
         (ii) In a pressureless medium the synchronous gauge is effectively
         the same as the comoving gauge ($v \equiv 0$).
         Evaluating Eq. (\ref{entropic}) in the comoving gauge, and 
         using Eq. (\ref{Poisson}) we can show 
         $\delta T_v/T = {1 \over 3} \delta_v 
         = {2 \over 9} (k/aH)^2 \varphi_\chi$ 
         at $E$, and thus neglible compared with $\varphi_\chi$ at $E$.
\end{thebibliography}
\end{document}